\documentclass[conference]{IEEEtran}
\usepackage{float}

\IEEEoverridecommandlockouts
\usepackage{amsmath,amssymb,amsfonts}
\usepackage{algorithmic}
\usepackage{authblk}

\usepackage{graphicx}
\usepackage[
backend=biber,
sorting=none,
url=false,
]{biblatex}
\usepackage{textcomp}
\usepackage{xcolor}
\AtNextBibliography{\footnotesize}
\def\BibTeX{{\rm B\kern-.05em{\sc i\kern-.025em b}\kern-.08em
    T\kern-.1667em\lower.7ex\hbox{E}\kern-.125emX}}
\begin{document}

\title{5-axis Multi-material Desktop Additive Manufacturing of Conformal Antennas}
\author{
 Iván Revenga Riesco}
 \author{
 Borut Lampret}
 \author{
 Connor Myant}
 \author{
 David Boyle}

 \affil{Imperial College London}

\maketitle

\begin{abstract}
This paper describes the novel use of low-cost, 5-axis, multi-material additive manufacturing to fabricate functional, complex conformal antennas. Using a customised open source 5-axis desktop printer incorporating conductive filaments, conformal S-band patch and Ultra-Wide Band antennas were fabricated and compared against planar-printed counterparts and electromagnetic simulations. Results show the potential of the approach for superior impedance matching, reduced fabrication time, and cost savings; highlighting the applicability of multi-axis multi-material prototyping of antennas with complex geometries.\\
\end{abstract}


\section{Introduction}
\begin{figure}[b]
    \centering
    \includegraphics[width=0.9\linewidth]{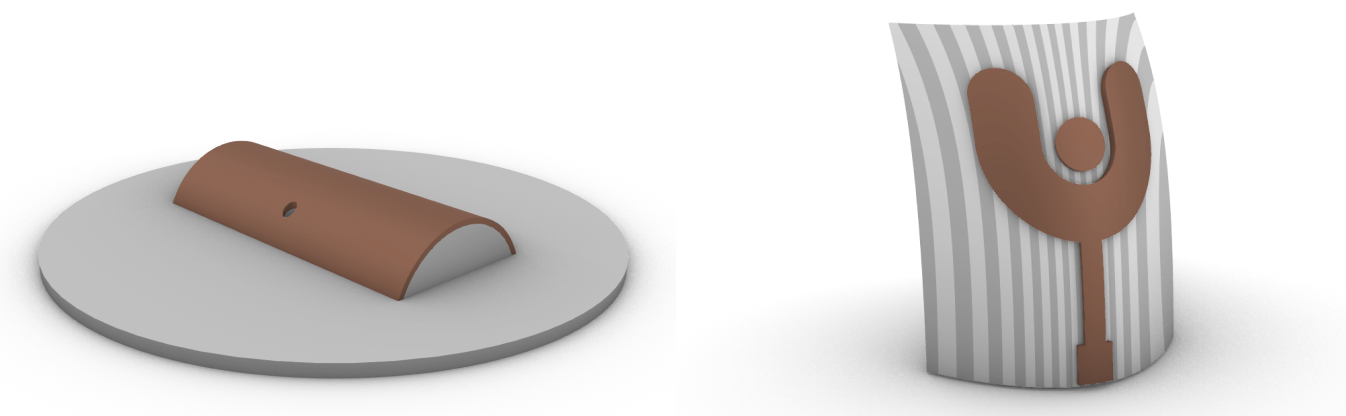}
    \caption{3D renders of conformal S-band patch antenna (left) and double-curvature UWB antenna (right)}
    \label{fig:3Dmodels}
\end{figure}

Additive Manufacturing (AM) is revolutionising the fabrication of antennas by enabling rapid prototyping and the creation of intricate geometries at a reduced cost \cite{helena_antenna_2020}. Traditional 3D printing methods for antennas, predominantly planar techniques, often face limitations due to added impedance at the layer interfaces along the z-axis \cite{hong_5-axis_2023}. This work explores the application of multi-axis material extrusion for the fabrication of conformal antennas, considering that improved deposition control enabled by multi-axis platforms may enhance performance and manufacturing efficiency. Two different conformal antenna designs were investigated for this purpose, pictured in Figure~\ref{fig:3Dmodels}. These include a conformal S-band patch antenna on a semi-circular substrate and a double-curvature Ultra-Wide-Band (UWB) antenna. The aim is to validate the potential of 5-axis multi-material desktop 3D printing by comparing the performance of samples manufactured in this way versus their planar-fabricated counterparts and respective electromagnetic simulations.

\section{Related work}
Additive manufacturing is gaining attention in antenna design due to its potential for cost reduction and design flexibility~\cite{helena_antenna_2020}. Techniques like VAT photopolymerisation have been used to produce antennas with complex geometries, including plated resin horn antennas and dielectric lenses \cite{ruihua_sla_2022,zhang_dielectric_2019}. For instance, PolyJet methods have enabled high-density out-of-plane antenna structures \cite{mohd_ghazali_3d_2019}, and hybrid approaches combining PolyJet with material extrusion have been used for phased array antennas \cite{shin_polymer-based_2019}.

Material extrusion (ME) has gained prominence due to its accessibility and suitability for low-cost fabrication. Conductive filaments such as Electrifi have been pivotal in fabricating functional 3D printed Radio Frequency (RF) circuits and antennas, including patch and Yagi-Uda antennas  \cite{fougeroux_performance_2024,colella_yagi-uda_2020}. However, when printing these conductive filaments, planar ME systems suffer from conductivity limitations along the z-axis \cite{hong_5-axis_2023}, making it difficult to harness the full potential of additive technology and create 3D circuits and electromagnetics. Recent advancements, like the open-source 5-axis printing ecosystem Open5x~\cite{hong_open5x_2022}, have made conformal printing more accessible, democratising access to this technology and demonstrating the benefits and challenges associated with it, including making novel electronic components~\cite{HONG2023103546}. This work builds on these developments, providing what we believe to be the first systematic exploration of 5-axis multi-material desktop printing of conformal antennas.

\section{Methodology}
\subsection{Antenna Design and Simulation}
Parametric models of the two antenna designs were developed in Ansys HFSS and Rhino 3D. The original patch antenna design was tuned to operate at 3~GHz \cite{muntoni_curved_2020}. The UWB antenna, following~\cite{balderas_low-profile_2019}, was designed for 3–16~GHz, but instead we projected it onto a double-curved substrate to demonstrate a more complex use case. We performed electromagnetic simulations that included frequency-dependent material models for PLA and Electrifi. These were built using data gathered from various experiments in the published literature \cite{colella_electromagnetic_2022}, and allowed the designs to be further optimised/tuned using these as the substrate and radiating element materials.

\subsection{Fabrication Workflow}
\begin{figure}[t]
    \centering
    \includegraphics[width=0.7\linewidth]{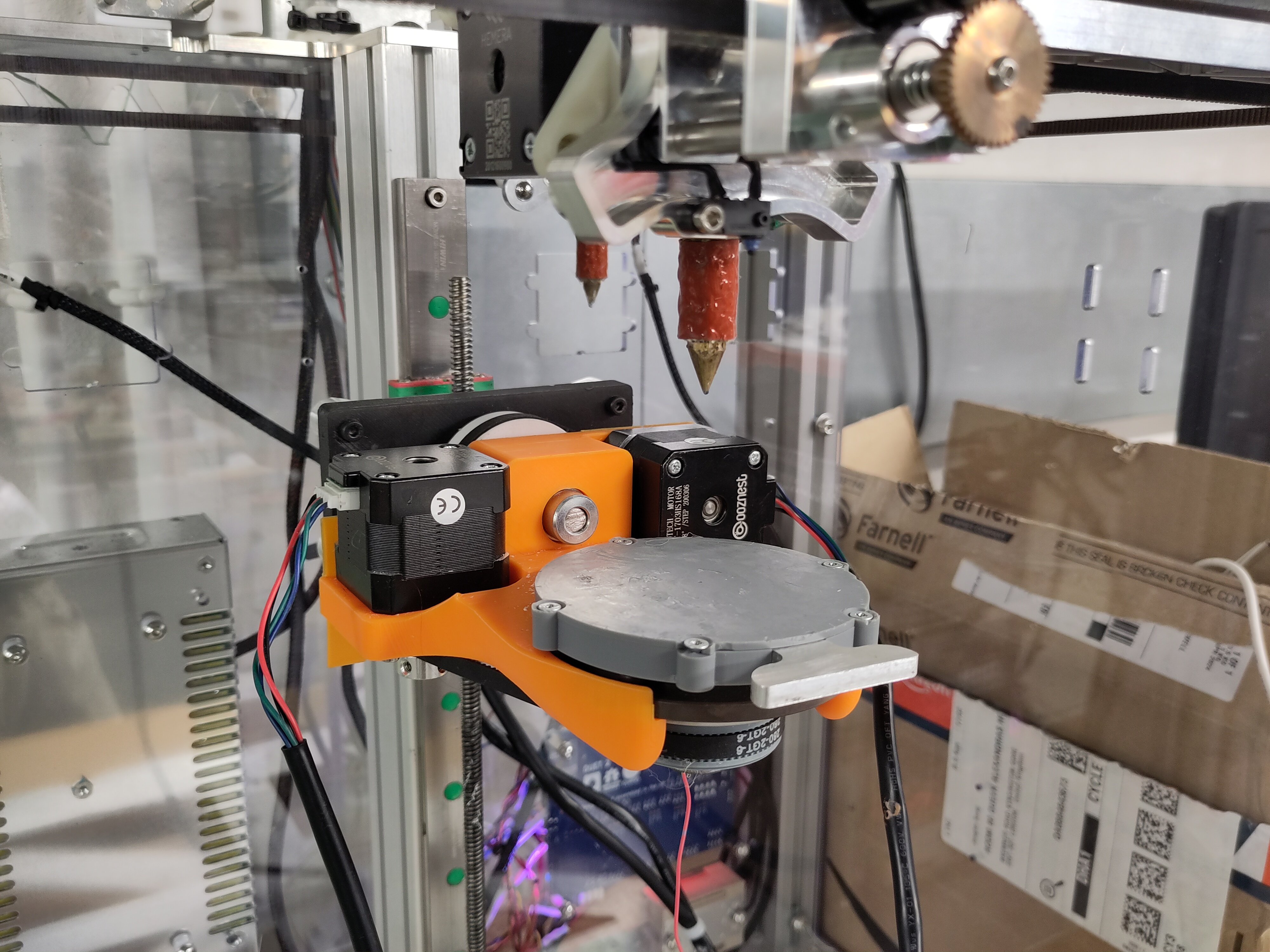}
    \caption{Open5x platform on an E3D tool-changer system}
    \label{fig:Open5x}
\end{figure}

Antenna samples were fabricated on a modified E3D tool-changer system with an Open5x platform for multi-axis printing, featuring long 0.4~mm nozzles~(Fig.~\ref{fig:Open5x})\footnote{YouTube video demo: https://youtu.be/AR7i1UPdsgc}. Tool-head 1 printed PLA (225~°C and 40~mm/s) and Tool-head 0 printed Electrifi (145~°C and 5~mm/s). The planar samples were printed conventionally using a dual-extruder configuration from PrusaSlicer. The multi-axis samples were printed using the Open5x Rhino-based path and G-code generators \cite{hong_open5x_2022}. Post-processing steps, including the addition of ground planes and SMA connectors, followed standard practices like the use of copper tape and silver epoxies \cite{fougeroux_performance_2024}. A total of 8 samples were produced, 2 for each design and fabrication approach combination. 

\subsection{Testing and Characterisation}
The dimensions of the fabricated samples were measured and compared among fabrication methods. To asses antenna performance, the S11 parameters were measured using a  MS2038C vector network analyser. The return loss and location of the resonant modes were compared among the samples and against simulation data. Lastly, the functionality of the UWB antenna samples was demonstrated using commercial UWB ranging nodes.
\section{Results and Discussion}

\subsection{Conformal Patch Antenna}
\begin{figure}[t]
    \centering
    \includegraphics[width=0.8\linewidth]{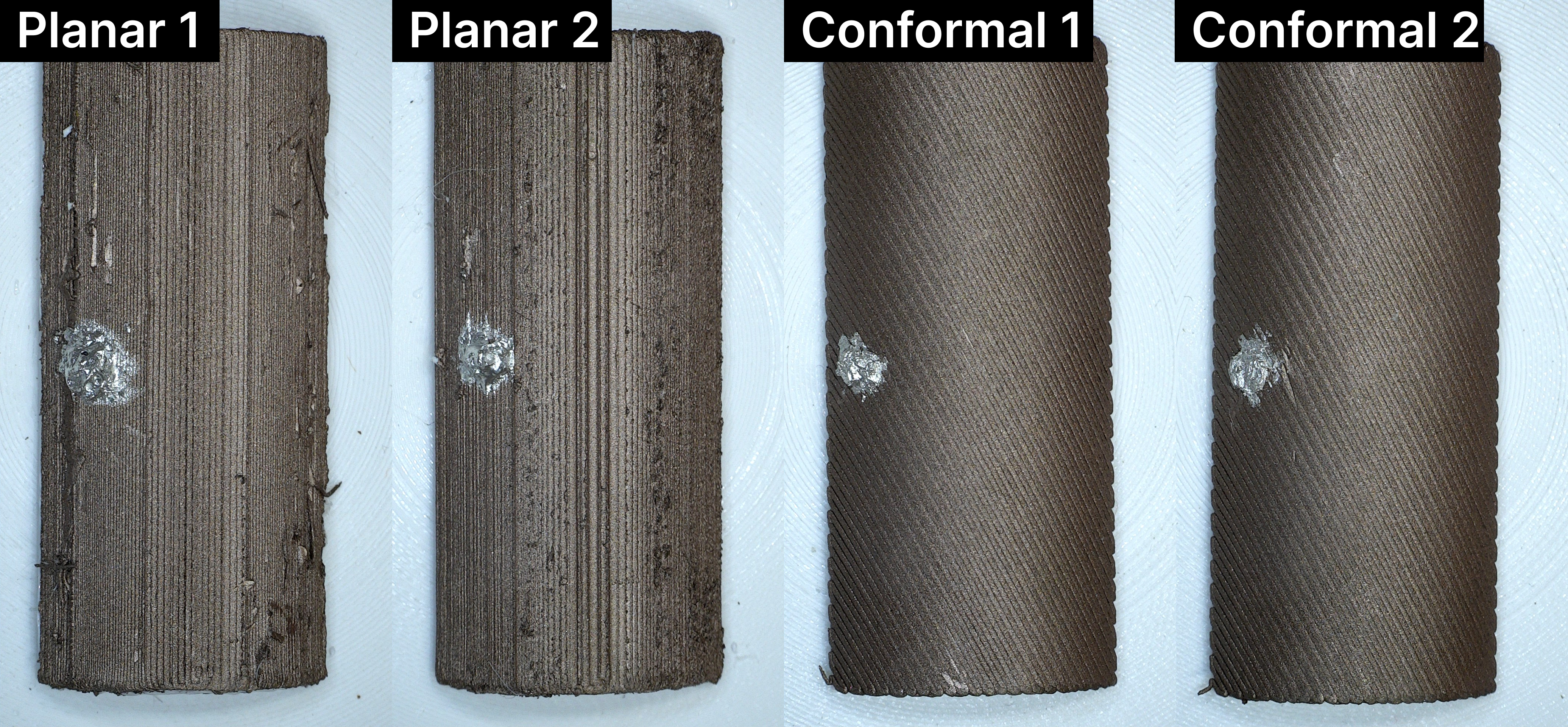}
    \caption{Comparison of sample patch antennas under microscope}
    \label{fig:patchall}
\end{figure}

\begin{figure}[t]
    \centering
    \includegraphics[width=0.7\linewidth]{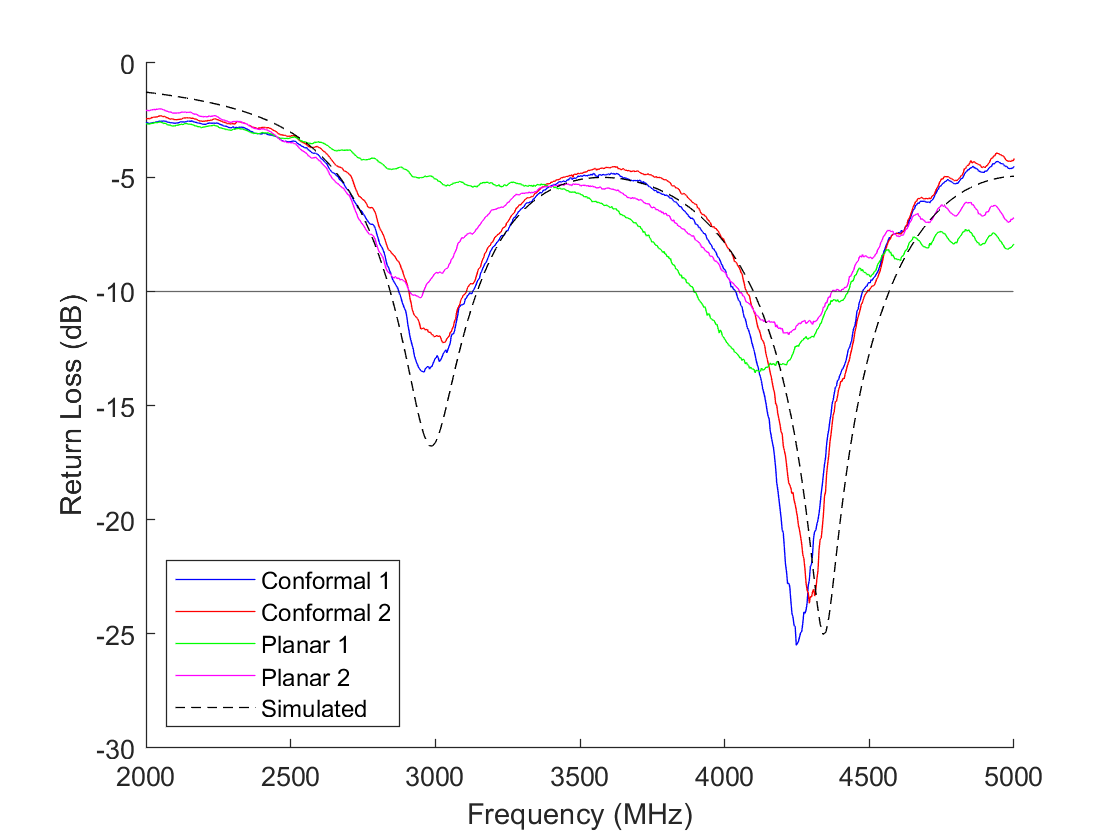}
    \caption{S11 parameters of simulated, planar-printed and multi-axis-printed (conformal) patch antenna samples}
    \label{fig:patchs11}
\end{figure}

The multi-axis samples exhibited superior build quality with smoother surfaces and fewer defects compared to planar samples (Fig.~\ref{fig:patchall}). Fabrication time and material usage were reduced by 28\% (3~hours and 30~minutes down to 2~hours and 30~minutes) and 10\% (19.7~g down to 17.8~g), respectively. Compared to the UWB antennas, the patch antennas had a significantly lower average dimensional error in the order of 2\% with feature size errors of maximum 12\% (Fig.~\ref{fig:dimerror}). Return loss measurements showed that multi-axis samples more closely matched simulated results (Fig.~\ref{fig:patchs11}), with reduced impedance mismatches at both resonant modes (3~GHz and 4.4~GHz), most notably at the second order mode with a difference of roughly 10~dB. By correlating the current flow in the resonant modes and different conductive trace depositions, it is suggested that anisotropic conductivity in 3D-printed conductive traces is the cause of the large discrepancies between the samples. This was further validated by performing electromagnetic simulations using anisotropic material models.

\subsection{Double-Curvature UWB Antenna}
\begin{figure}[t!]
    \centering
    \includegraphics[width=0.8\linewidth]{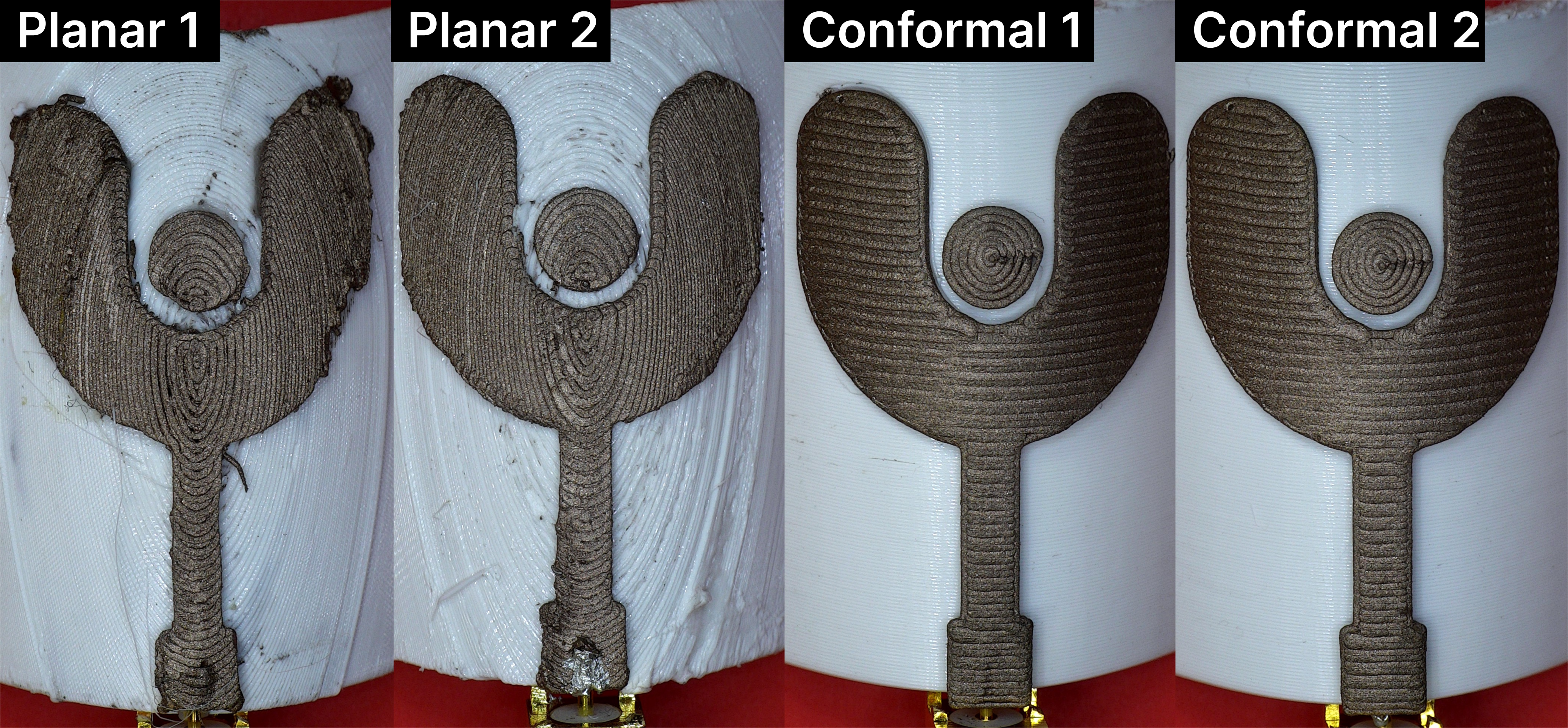}
    \caption{Comparison of sample UWB antennas under microscope}
    \label{fig:uwball}
\end{figure}

\begin{figure}[t]
    \centering
    \includegraphics[width=0.7\linewidth]{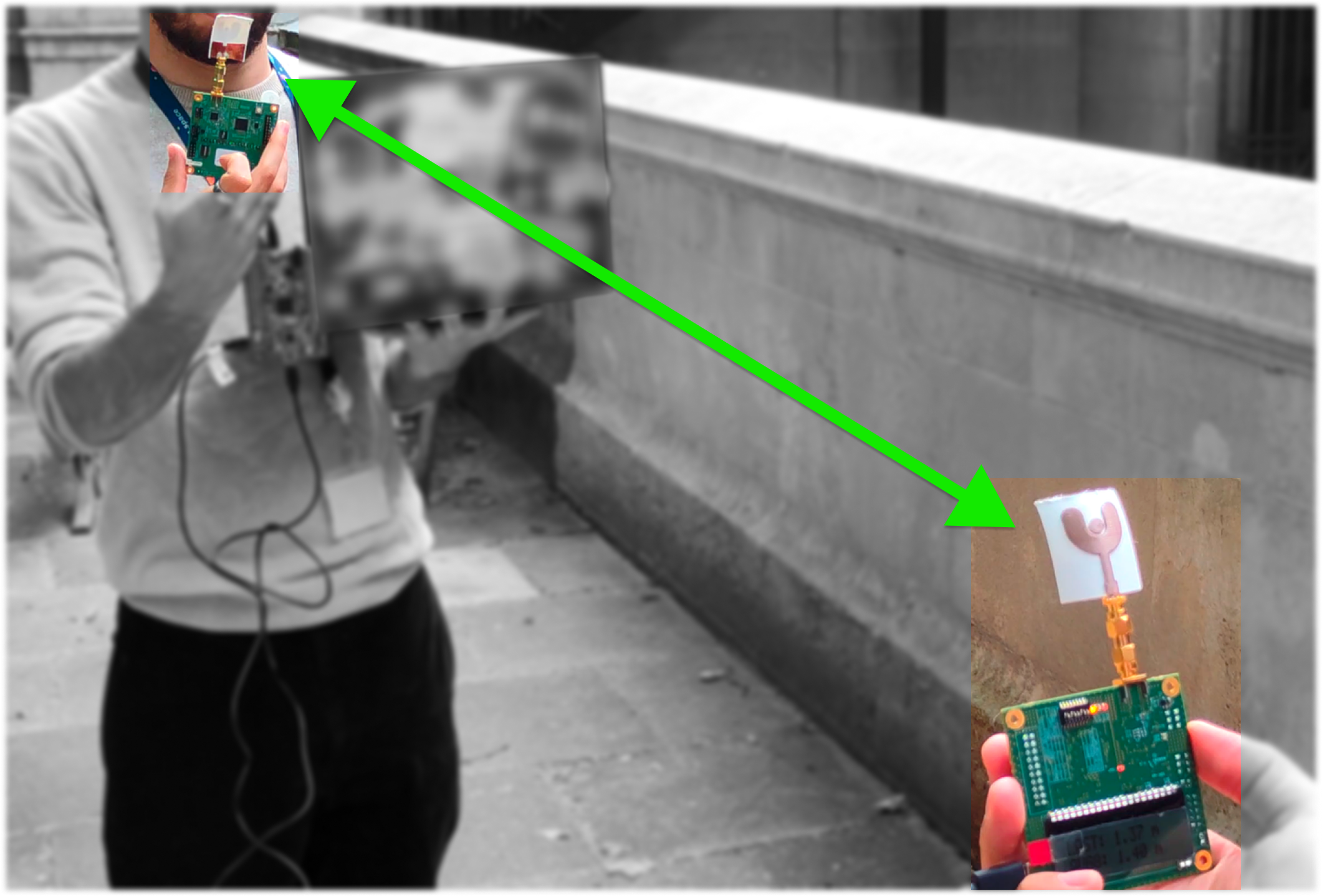}
    \caption{Conformal UWB antenna functional test with DW1000}
    \label{fig:demo}
\end{figure}

\begin{figure}[t]
    \centering
    \includegraphics[width=0.8\linewidth]{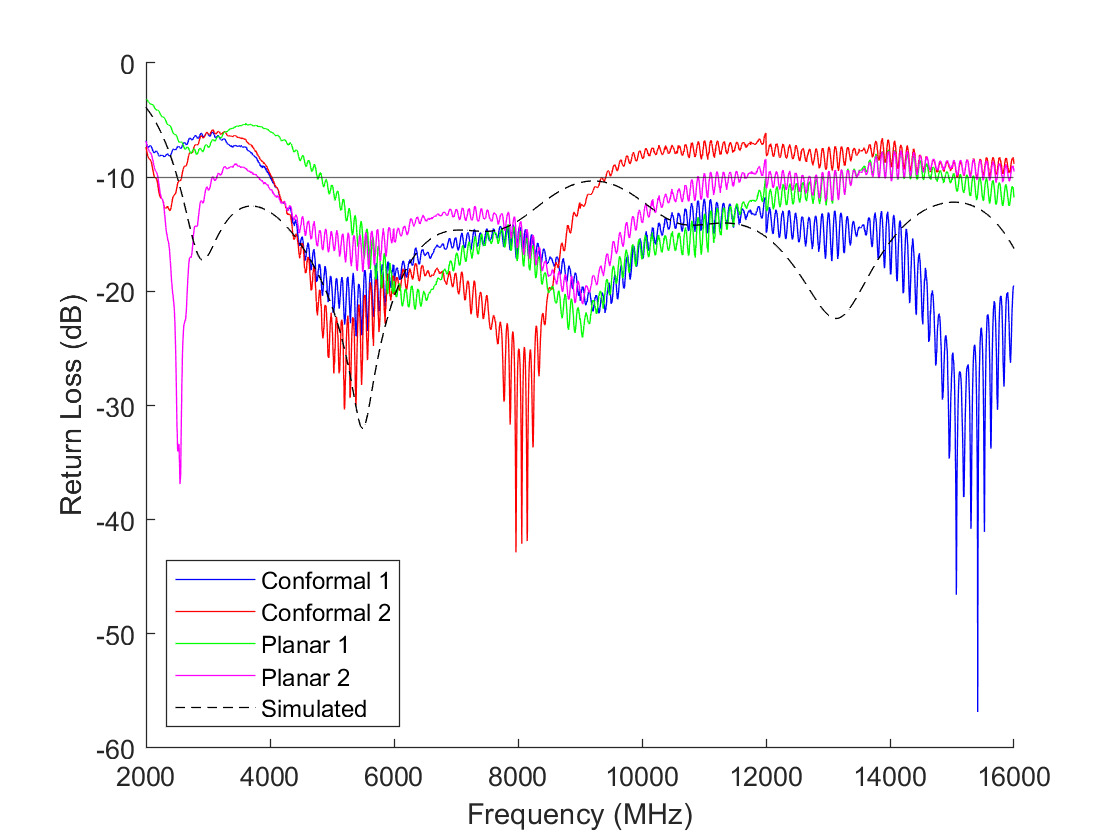}
    \caption{S11 parameters of simulated, planar-printed and multi-axis-printed (conformal) UWB antenna samples}
    \label{fig:uwbs11}
\end{figure}

\begin{figure}[h]
    \centering
    \includegraphics[width=.95\linewidth]{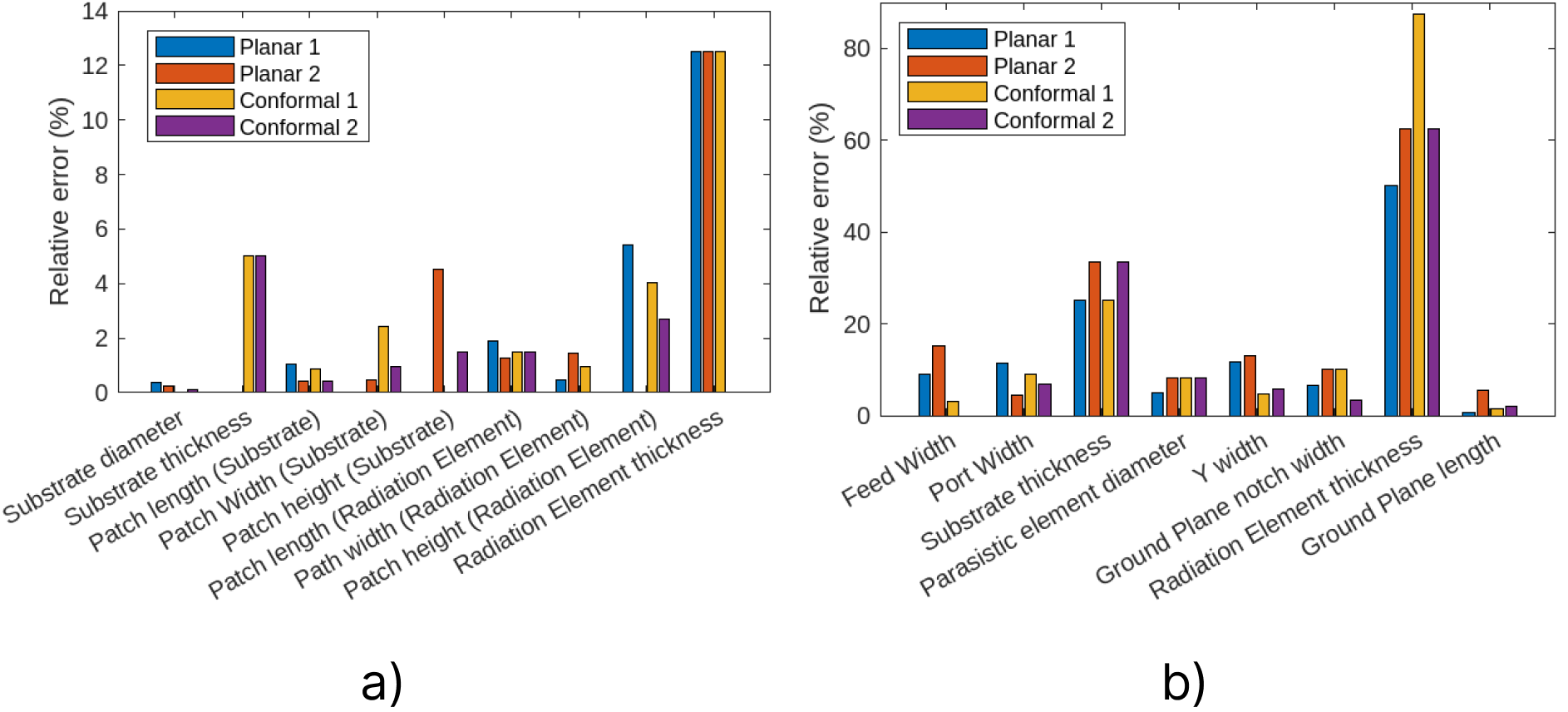}
    \caption{Relative error comparison among samples and antenna designs for the patch antennas (left) and UWB antennas (right)}
    \vspace{-10pt}
    \label{fig:dimerror}
    
\end{figure}

For the complex UWB design, the multi-axis method had better surface finish (Fig.~\ref{fig:uwball}), reduced print time by 50\% (2~hours and 15~minutes down to 1~hour and 10~minutes) and the material usage by 43\% (5.9~g down to 3.4~g), demonstrating significant manufacturing efficiency gains. These samples had an average dimensional error of about 15\% with some features even surging above 80\% relative error (Fig.~\ref{fig:dimerror}). In terms of return loss measurements, all of the fabricated antennas exhibited large and non-trivial deviations from the simulation (Fig.~\ref{fig:uwbs11}). This is thought to be due to the sensitivity of higher frequency (+6~GHz) designs and manufacturing tolerances being too similar in magnitude and general repeatability issues. Despite this, most of the samples show good return loss and operational bandwidth, demonstrating the ability of some samples to operate throughout most IEEE-defined UWB channels \cite{molisch_ultra-wide-band_2009}. Lastly, functional tests using DW1000 UWB hardware were successfully carried out, showing effective integration with COTS systems~(Fig.~\ref{fig:demo})\footnote{YouTube video demo: https://youtu.be/Ih8vxaL6uqU}.

\subsection{Implications of Multi-Axis Fabrication}
Multi-axis printing offers distinct advantages for conformal antenna fabrication, including less printing time, material use and better impedance matching, allowing for more efficient design iteration. Manufacturing advantages become more prominent as the complexity increases, whereas the higher deposition control is thought to provide more design freedom and result in better impedance matching. Greater design freedom allows antenna designers to have greater control over the inherent anisotropic effects of additive manufacturing.  With such control, designers can also attempt to leverage anisotropy, e.g., purposefully minimising certain resonant modes. 

\section{Conclusion}
This work demonstrates the potential of low-cost, 5-axis, multi-material additive manufacturing for the production of complex conformal antennas. The approach provides improved fabrication efficiency and enhanced impedance matching resulting in the ability to produce faster and better prototypes. Future research directions could focussing on further characterisation of anisotropic behaviours, material property variability of the conductive filaments available, performance consistency of 3D printed RF elements, and incorporating these practical considerations within antenna design and simulation software.

\section{Acknowledgements}
The authors thank Freddie Hong for insights on 5-axis printing, Nunzio Pucci, Peilong Feng and Oleksiy Sydoruk for their help with characterisation of the antenna samples, and UKRI EPSRC for financial support (EP/Y037421/1).

\begin{tiny}
\printbibliography
\end{tiny}

\end{document}